\documentclass[prl,superscriptaddress,showpacs,twocolumn]{revtex4-1}
\usepackage{graphicx,amsmath,amssymb,natbib}
\begin{document}
\title{Orientational order of carbon nanotube guests\\
 in a nematic host suspension of colloidal viral rods}

\author{Nicolas Puech,$^1$ Matthew Dennison,$^2$ Christophe Blanc,$^3$ Paul van der Schoot,$^{4,5}$ Marjolein Dijkstra,$^2$ Ren\'e van Roij,$^5$ Philippe Poulin,$^1$ and Eric Grelet}
\altaffiliation[E-mail: ]{grelet@crpp-bordeaux.cnrs.fr}
\affiliation{Universit\'{e} de
Bordeaux, Centre de Recherche Paul-Pascal - CNRS, 115 Avenue Schweitzer, 33600 Pessac, France\\
$^2$ Debye Institute for Nanomaterials Science, Utrecht University, Princetonplein 5, 3584 CC Utrecht, Netherlands\\
$^3$ Universit\'{e} Montpellier 2 et CNRS, Laboratoire Charles Coulomb, Place E. Bataillon, 34095
Montpellier, France\\
$^4$ Faculteit Technische Natuurkunde, Technische
Universiteit Eindhoven, Postbus 513, 5600 MB Eindhoven, Netherlands\\
$^5$ Instituut voor Theoretische Fysica, Universiteit Utrecht,
Leuvenlaan 4, 3584 CE Utrecht, Netherlands}

\date{\today}

\begin{abstract}
In order to investigate the coupling between the degrees of alignment of elongated particles in binary nematic dispersions, surfactant stabilized single-wall carbon nanotubes (CNTs) have been added to nematic suspensions of colloidal rod-like viruses in aqueous solution. We have \textit{independently} measured
the orientational order parameter of both components of the guest-host system by means of polarized Raman spectroscopy 
and by optical birefringence, respectively. 
Our system allows therefore to probe the regime where the guest particles (CNTs) are  shorter and thinner than the fd virus host particles. We show that the degree of order of the CNTs is systematically smaller than that of the fd virus particles for the whole nematic range. These measurements are in good agreement with predictions of an Onsager-type second-viral theory, which explicitly includes the 
flexibility of the virus particles, and the 
polydispersity of the CNTs.

\end{abstract}


 \maketitle

The alignment of colloidal particles in liquid crystals has been the topic of
intensive study in recent years \cite{Andrienko02,Burylov94,Hung09,Lapointe04,Weiss06,Mondiot09,Dogic04,Tkalec08,Schoot08,Dierking05,
Jeong07,Lagerwall08,Lynch02,Scalia08,Shah08}. Apart from scientific interest, motivation for such
studies is found in the potential applications of oriented functional
particles such as carbon nanotubes, metallic or
semi-conducting nanowires and nanoribbons. Typically, the size of the
colloidal inclusions is much greater than the size of the nematogens,
i.e., the molecular building blocks of the liquid crystal. Indeed, in most
common thermotropic and lyotropic liquid crystals, the nematogens are small molecules.
The nematic medium can therefore be considered as
a continuous matrix in which colloidal inclusions are embedded. Under
these conditions, the behavior of the liquid crystal is well
described by a phenomenological approach based on continuum
elasticity, interfacial energy and surface tension
anisotropy. Anisotropic colloidal inclusions in conventional thermotropic nematic
liquid crystals orient in response to elastic torque and surface
tension anisotropy \cite{Andrienko02,Burylov94,Hung09,Lapointe04,Mondiot09,Tkalec08}.
These effects are generally very strong with
associated energies vastly exceeding the thermal energy $k_BT$ that
drives Brownian rotation. As a consequence anisotropic colloidal
particles exhibit a degree of orientation which is greater than that of
the nematogens in which they are dispersed.

Recent developments involving novel colloids and nanoparticles,
including nanotubes, nanowires, nanoribbons, and liquid crystals made
of particles in the colloidal size domain, raise new questions related to the ordering of
colloids in liquid crystals in which the nematogens are of a size that
compares with or exceeds the size of the inclusions. Here, we report on an
exploration of this new regime, by studying the case of single-wall carbon nanotubes
(CNTs) embedded in a lyotropic colloidal nematic liquid crystal of fd virus particles in water.
The fd virus can be seen as a semi-flexible
polyelectrolyte with a contour length of $L_{fd}=0.88~\mu$m, a bare diameter
of $D_{fd}=6.6$~nm and a persistence length of about
$P_{fd}=2.2~\mu$m \cite{Dogic06}. The liquid-crystalline phases of
fd virus suspensions have been studied extensively, because the fd viruses
are usually considered due to their monodispersity as an ideal rod
system for comparison with theory \cite{Dogic06,Grelet08,Barry09,Pouget11,theory1,theory2}.
CNTs with an average length of about 0.3~$\mu$m \cite{CNT3,Puech11}, therefore shorter than the fd virus particle,
are added at a small concentration to the fd virus nematic phase.
The CNT colloidal stability is provided by the adsorption of suitable
surfactant molecules (in our case a bile salt), which results in an overall diameter of
about 2~nm for the surfactant stabilized carbon nanotubes \cite{CNT3,Puech11}.
Our mixture of fd virus and CNT particles of different lengths is reminiscent of the
bi-disperse suspensions of rigid rod-like particles studied theoretically by Lekkerkerker and collaborators
\cite{Lekkerkerker84}. Nevertheless experimental characterization of orientational
parameters in conventional bi-disperse systems made of similar particles is not straightforward
since physical properties of the particles do not depend on their
size. As a consequence some key features such as distinct degree of ordering of the particles can not be determined.

\begin{figure}
    \includegraphics[width=0.66\textwidth,angle=270]{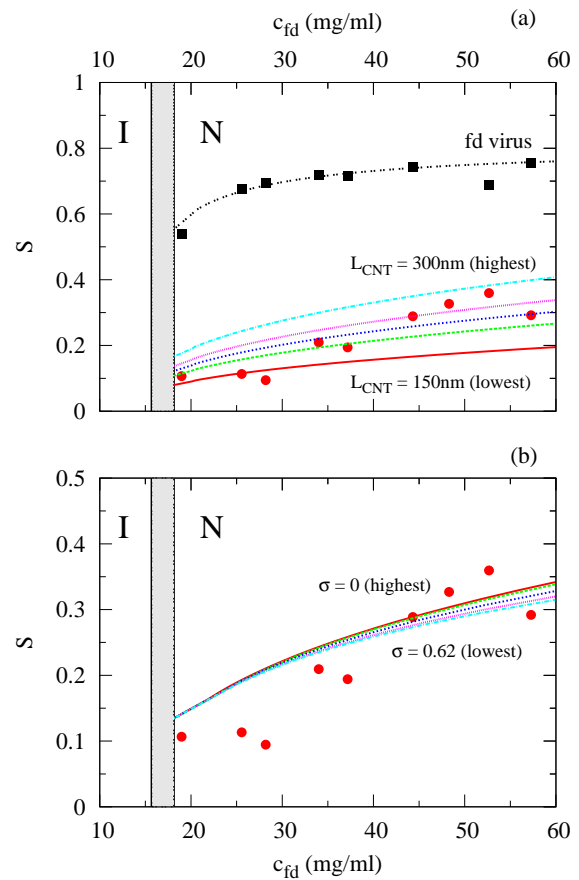}
    \caption{\label{fig1} Measured (symbols) and calculated (lines)
     orientational order parameters of the
      host nematic phase of fd virus suspensions $S_{fd}$ (black
      squares) doped with surfactant-stabilized single wall carbon nanotubes
      $S_{CNT}$ (red circles) as a function of the virus concentration
      $c_{fd}$. Theoretical results are for: (a) monodisperse ($\sigma=0$) CNTs with different lengths $L_{CNT}$;
      (b) polydisperse CNTs with fixed average length $\langle L_{CNT} \rangle=250\mathrm{nm}$ and a variety
      of length polydispersities $\sigma$.
      The different lengths in (a) are $L_{CNT}$ $150\mathrm{nm}$
      (lowest line), $200\mathrm{nm}$, $225\mathrm{nm}$,
      $250\mathrm{nm}$ and $300\mathrm{nm}$ (highest line). The different CNT polydispersities in (b) are $\sigma=0$ (highest line),
      0.13, 0.36, 0.52 and 0.62 (lowest line). The two vertical lines denote the isotropic liquid (I) and nematic (N) binodals.
      }
  \end{figure}

Our system of choice is quite a unique model system, because it turns out
that the orientational order parameter of the two components (CNTs and fd
viruses) 
 can be independently measured. The order parameter of CNTs can be probed by
polarized Raman and photoluminescence spectroscopies \cite{Puech11,Zamora09,Saito11},
 and that of fd virus particles by optical birefringence \cite{Purdy03,Kang07}.
Indeed, resonant Raman scattering of CNTs and their polarization
dependent response offer the
opportunity to measure the CNT mean orientation at low concentrations down
to the limit in which the systems do not phase separate and the
inclusions do not appreciably affect the ordering of the host particles (dilute limit). The
distinctive features of our system allow us to experimentally
determine the order parameter of the CNTs as a function of the order
parameter of the host liquid crystal, the latter being set by the concentration of the viruses.
This experimental progress offers
thereby an opportunity for comparisons with theories that could not be
tested with other systems. Our finding that the orientational
ordering of CNT qualitatively increases with that of the fd virus is expected.
In contrast to earlier results for rod-like guest colloids dispersed in low molecular-weight liquid crystals, 
we show here that the degree of orientational order of the guest particles is now smaller than that of the host.
Counter-intuitively, this distinctive feature suggests that using host rods with increasing aspect ratio to better orient guest nanorod particles may not lead to better alignment. In addition to providing useful guidance for the controlled processing of functional composites, the present result offers a unique route towards the development of composites that combine a high degree of alignment of the matrix components while preserving some disorder of the smaller inclusions. Such a combination is of great technological interest for composites that are expected to exhibit good mechanical properties arising from the alignment of the matrix components and electrical or thermal conductivity arising from the formation of percolated networks of the inclusions. Usually a strong alignment of the guest functional nanorods results in a loss of contact probability and thereby in a loss of electrical conductivity.
In this work, the order parameter of the guest CNTs increases from 0.1 to
0.35 as the order parameter of the host viruses is increased from 0.55 to
0.75 (Fig.~\ref{fig1}(a)). These results are quantitatively described by a theoretical model,
which includes both the flexibility of the virus and the CNT polydispersity in length.

Experimentally, the CNT dispersions were initially prepared from an aqueous suspension
of single-wall carbon nanotube bundles (furnished by Elicarb batch K3778) and
dispersed by bile salt surfactant (an equimolar mixture of sodium cholate
and sodium deoxycholate), at the respective concentrations of 0.5\%
w/w CNTs and 0.5\% w/w bile salts. To exfoliate the CNT bundles,
sonication was applied to the suspension for a period of three
hours. A purification process by selective centrifugations was subsequently
performed on the nanotube suspensions. After removing nanotube aggregates by
centrifugation at low speed (30 min, 2000 $g$), 
the longest carbon
nanotubes exhibiting some entanglements and structural defects were removed by
two ultracentrifugation steps (45 min at 200000 $g$ 
and 180 min at 200000 $g$, 
respectively). Finally, a purified surfactant stabilized CNT
suspension was obtained at a concentration of 0.34\% w/w, 
as measured by thermogravimetric analysis.
Sonication and centrifugation
allow a selection of short and straight particles, which can be really considered
as rigid rods, in contrast to raw and long nanotubes that exhibit a pronounced waviness or tortuosity \cite{Puech11}.

A batch of fd virus prepared by standard biological protocols and dialyzed against a
TRIS-HCl buffer at pH=8.2 \cite{Dogic06} was concentrated at $c_{fd}=$90~mg/mL
as determined by spectrophotometry i.e. close to the chiral nematic -
smectic phase transition \cite{Grelet08}. By starting with a dispersion composed
of 20\% w/w of bile salt stabilized CNT dispersion and 80\% w/w of fd
virus suspension, further samples were prepared by mass dilution with
buffer, keeping constant the fd virus-to-CNT ratio. 
The samples of varying dilutions were then placed
in quartz capillary tubes (0.7~mm in diameter). 
 They were aligned first with an NMR magnet producing a field strength of about 4.6 T
during 6 hours, and after that inserted in a smaller in-house made permanent
magnet of about 1.5 T allowing observations with optical microscopy and spectroscopy measurements by
maintaining the sample orientation. The director of the host nematic phase of virus particles aligns along the
applied magnetic field, resulting in a single uniform domain. Note that CNTs exhibit a very weak magnetic anisotropic, and cannot be significantly aligned only by using an external magnetic field \cite{Islam}.
We found the samples to be more difficult to align
for the highest concentrations in the nematic range, thereby yielding
a lower alignment quality and a lower measured orientational order parameter as
compared to reported values of pure fd virus suspensions \cite{Purdy03
}.

\begin{figure}
\includegraphics[width=0.45\textwidth]{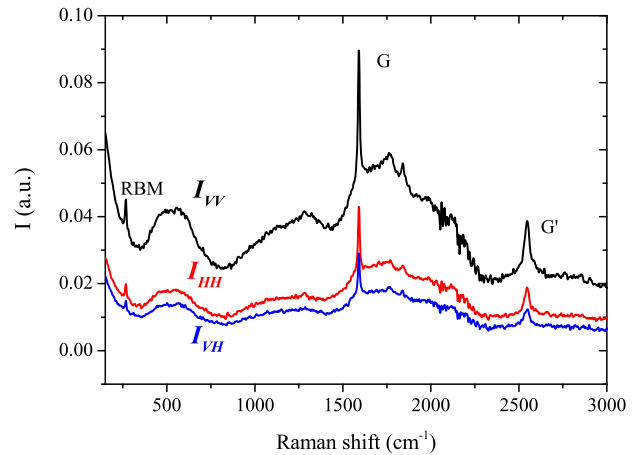}
\caption{\label{Raman} Polarized Raman and photoluminescence spectra
  of CNTs at 0.027\% w/w dispersed in a solution of fd virus at a concentration of
  $c_{fd}=$34 mg/mL. The orientational order parameter $S_{CNT}$ is
  obtained from the three polarization configurations $I_{VV}$,
  $I_{VH}$, and $I_{HH}$ using Eq. \ref{Scnt}. The RBM, G and G' Raman bands can be clearly
  distinguished.  }
\end{figure}

Polarized Raman and polarized
photoluminescence microspectroscopies
were used to
determine the CNT orientational order parameter, $S_{CNT}$. We worked with the 
1064 nm line of a Nd:YAG laser and a Fourier transform Bruker RFS100 spectrometer which detects in the 900-1700 nm range.
This setup allows the simultaneous measurements of the NIR Raman signal of all CNTs and the large NIR photoluminescence of individual tubes. The
presence and the shape of RBM (radial-breathing mode), G, and G' bands in the spectra (Fig. \ref{Raman}) are
consistent with the occurrence of mainly single-wall carbon nanotubes (double-wall ones
are also present in Elicarb samples but in a small proportion) \cite{Puech11,CNT5}. The existence of a strong photoluminescence in these
spectra with no time evolution indicates that a large fraction of CNTs
are individually well-dispersed in solution. The Raman spectra of an
oriented domain, shown in Fig. \ref{Raman}, are obtained in the three main
polarization configurations, $I_{VV}$, $I_{HH}$, and $I_{VH}$ where
the first and second subscripts respectively refer to the
polarizations of incident and scattered beams, oriented either
parallel V or perpendicular H to the alignment direction i.e., along
the magnetic field. 
The scattered intensity in the VV configuration
is much greater than the intensity for the two other ones, showing a
significant orientation of nanotubes. The orientational order parameter
is obtained from the intensity over all the whole studied range of
wavelengths by \cite{CNT5}:
\begin{equation}
S_{CNT}=\frac{3I_{VV}+3I_{VH}-4I_{HH}}{3I_{VV}+12I_{VH}+8I_{HH}}. \label{Scnt}
\end{equation}
Note that pure viral suspensions containing no carbon nanotubes do not exhibit any
signal in polarized Raman spectroscopy. The orientational order
parameter of fd virus, $S_{fd}$, has been independently determined by
optical birefringence ($\Delta n$) measurements, where the virus
director field coincides with the main optical axis. The degree of
alignment is given by \cite{Purdy03}:
\begin{equation}
S_{fd}=\frac{\Delta n/c_{fd}}{\Delta n_{sat}/c_0}. \label{Sfd}
\end{equation}
where $\Delta n_{sat}/c_0=3.8\times 10^{-5}$~mL/mg is the specific
normalized birefringence of perfectly aligned viral rods \cite{Purdy03}.
Such birefringence measurements have been performed on
aligned samples with the use of a Berek compensator. The very small
amount of CNTs (between 0.01 to 0.06 \% w/w) added to the suspension
of virus particles does not alter the optical properties of the
sample. Moreover the use of Berek compensator provides a direct visual
measurement of the induced birefringence, which is rather insensitive
to light scattering and dichroism artifacts. Indeed, the measurement does
not depend on the absolute transmitted light intensity, but only on
the position of the zero-order interference band minimum
\cite{Dozov11}. According to the isotropic liquid and nematic binodal
concentrations and taking into account all the charged species present in
solution, the ionic strength has been determined to be I$\simeq$30 mM \cite{SM}.
A set of experimental data consisting of both fd virus and CNT orientational
order parameters has thus been obtained, and is presented in Fig.~\ref{fig1}(a).

Our results are well described by using a modified Onsager
theory that models the guest-host system as a binary mixture of semiflexible (fd) and rigid
(CNT) colloidal rods interacting through a hard-core excluded volume interaction \cite{SM}.
The segmented chain method described in Refs.~\cite{theory1,theory2} was used to model the semi-flexible rods. We
  take component $1$ to be the bulk fd virus liquid crystal, with the
  fd rod parameters ($L_{fd}, D_{fd}, P_{fd}$) given above. As
  fd virus is charged, an effective diameter needs to be introduced in the
  calculations, which has been defined as the diameter that gives the same
  coexistence densities as those found experimentally, and this is
  known to depend on the ionic strength \cite{theory3}.
  Component $2$ is taken to be a distribution of CNTs that are stiff,
  polydisperse in length, with a fixed diameter of 2~nm \cite{SM}, and at a
  sufficiently low concentration so that they (i) do not interact with
  each other, and (ii) do not affect the density dependence of the
  fd virus ordering. Finally, a log-normal CNT length distribution has been used,
 given by:

  \begin{equation}
    \label{eq:ln}
    \Delta(l) = C l^{-1}\exp\left[-\frac{(\ln
        [l]-\mu)}{2\omega^{2}}^{2}\right],
  \end{equation}
  where $C$ is a normalization factor, such that $\int
  dl\Delta(l)=1$. The polydispersity is defined by
   $\sigma=\sqrt{\langle l^{2} \rangle / \langle l \rangle ^2-1}$, and can be expressed in terms of
   the mean $\mu$ and the standard deviation $\omega$ of $\ln[l]$. Here $l=L_{CNT}/L_{0}$
  with $L_{CNT}$ the CNT length and $L_{0}$ a reference length. For a
  polydisperse CNT system without fd virus, the choice of $L_{0}$
  is arbitrary provided sufficiently long rods \cite{theory4}. In the
  mixture considered here, however, a length-scale is set by the fd virus contour length $L_{fd}$,
 and in this case $L_{0}$ is chosen such that for a given $\sigma$ we fix the mean CNT length
 $\langle L_{CNT} \rangle$ defined by:

  \begin{equation}
    \label{eq:mean_l}
    \langle L_{CNT} \rangle = L_{0}\displaystyle \int dl \Delta(l) l.
  \end{equation}
  The CNT nematic order parameter is then given by:

  \begin{equation}
    \label{eq:mean_l}
    S_{CNT} = \displaystyle \int dl \Delta(l) S_{CNT}(l).
  \end{equation}
where $S_{CNT}(l)$ is the order parameter of a CNT of reduced
length $l$. This depends on the fd virus concentration through the orientational distribution function,
and is calculated using the method given in Refs. \cite{theory1,theory2}.
The CNT length distribution has been discretized, considering $180$ lengths
ranging from $L_{CNT}=50~\mathrm{nm}$ to $1000~\mathrm{nm}$, for which we
found that our results have converged to the continuum limit.

  In order to characterize the CNTs, we consider two  cases. Firstly, $S_{CNT}$ is calculated
  for a range of $\langle L_{CNT} \rangle$ values at $\sigma=0$, which
  corresponds to monodisperse CNTs dissolved in fd viruses, as a
  function of $c_{fd}$. Secondly, $S_{CNT}$ has been determined for a range
  of CNT polydispersities $\sigma$ as a function of the concentration of
  the background fd virus, $c_{fd}$. Both results are displayed in Fig.~\ref{fig1}, which indicates
   that the dependence of $S_{CNT}$ upon
  $\langle L_{CNT} \rangle$ is strong, whereas the results are surprisingly insensitive to the
 CNT polydispersity for a fixed mean CNT length
  $\langle L_{CNT} \rangle$. At low $c_{fd}$ just above the binodal, we observe that highly polydisperse
  CNTs with $\sigma$ as large as 62$\%$ exhibit essentially the same ordering as monodisperse ($\sigma=0$) CNTs
  with the same average length. At higher fd concentrations, the difference between polydisperse and monodisperse
  CNTs is bigger, but remains marginal. We thus reach the unexpected conclusion that the key parameter to
  account for the CNT ordering in the host fd suspension is
  the {\it average} CNT length, and {\it not} their length polydispersity.
  Clearly, this finding could largely simplify future modeling of such asymmetric hybrid systems.
 From our theoretical analysis, it can be also extracted that the present system is best described by
 $\langle L_{CNT} \rangle \sim
  200-250~\mathrm{nm}$, which is very consistent with the one expected for carbon nanotubes cut
  and sorted by sonication and centrifugation \cite{Puech11,CNT3}.

In summary, we have investigated experimentally and theoretically the coupling between the orientational order parameters of guest-host systems, in which guest carbon nanotubes are dispersed in the host nematic phase of fd virus. We focused attention on guest particles that are smaller in size than the host nematogens, preventing the description of the nematic phase as a continuum. We find that the orientational order parameter of the guest particles is lower than that of the host particles for all the studied concentrations.
This implies that using probes to measure the degree of order in nematics only gives qualitative - but not quantitative - information \cite{Lettinga00}. 
The present results can pave the way in designing and modeling
new hybrid materials with ordered nanosized and rod-like
particles. In particular it conceptually provides a unique route towards the development of composites that combine mechanical strength and electrical conductivity. This combination is challenging since most available approaches to align the matrix components of a composite (such as drawing, flow induced alignment, or alignment in an external field) lead to a strong increase of the orientation of the embedded functional nanorods, 
being then inefficient at providing composites with both good mechanical and electrical properties. The approach developed here offers a new possibility where the main matrix components remain strongly aligned and serve as reinforcements in composites while the embedded guest nanorods preserve some disorientation as requested to achieve a high density of electrical contacts yielding a good conductivity.

\begin{acknowledgments}
This work was financially supported by ANR, FOM and NWO-VICI grants. We thank E. Anglaret for fruitful discussions and for the use of FTIR Raman microscope.
\end{acknowledgments}

\end{document}